# Anfrage-getriebener Wissenstransfer zur Unterstützung von Datenanalysten

Andreas M. Wahl, Gregor Endler, Peter K. Schwab, Sebastian Herbst, Richard Lenz

Friedrich-Alexander-Universität Erlangen-Nürnberg
Lehrstuhl für Informatik 6 (Datenmanagement)
{andreas.wahl | gregor.endler | peter.schwab | sebastian.herbst | richard.lenz}@fau.de

**Abstract:** In größeren Organisationen arbeiten verschiedene Gruppen von Datenanalysten mit unterschiedlichen Datenquellen, um analytische Fragestellungen zu beantworten. Das Formulieren effektiver analytischer Anfragen setzt voraus, dass die Datenanalysten profundes Wissen über die Existenz, Semantik und Verwendungskontexte relevanter Datenquellen besitzen. Derartiges Wissen wird informell innerhalb einzelner Gruppen von Datenanalysten geteilt, jedoch meist nicht in formalisierter Form für andere verfügbar gemacht. Mögliche Synergien bleiben somit ungenutzt.
Wir stellen einen neuartigen Ansatz vor, der existierende Datenmanagementsysteme mit zusätzlichen Fähigkeiten für diesen Wissenstransfer erweitert. Unser Ansatz fördert die Kollaboration zwischen Datenanalysten, ohne dabei etablierte Analyseprozesse zu stören. Im Gegensatz zu bisherigen Forschungsansätzen werden die Analysten beim Transfer des in analytischen Anfragen enthaltenen Wissens unterstützt. Relevantes Wissen wird aus dem Anfrageprotokoll extrahiert, um das Auffinden von Datenquellen und die inkrementelle Datenintegration zu erleichtern. Extrahiertes Wissen wird formalisiert und zum Anfragezeitpunkt bereitgestellt.

## 1 Einführung

Zur Beantwortung von analytischen Fragestellungen sind meist mehrere heterogene Datenquellen erforderlich. Die benötigten Daten sind oftmals semi-strukturiert und liegen heutzutage nicht mehr ausschließlich in relationaler Form vor, sondern werden in den verschiedensten Formaten und Systemen gespeichert. So fallen beispielsweise bei der Patientenüberwachung in der Intensivmedizin neben strukturierten Patientenstammdaten auch große Mengen semi-strukturierter Medikationspläne, Datenströme von Vitalparametern oder Freitextnotizen an [Sa11]. Um solche Daten, unter anderem für die klinische Forschung, effizient auswertbar zu machen, können neuartige Ansätze wie die von Stonebraker et al. vorgeschlagenen Polystores verwendet werden [St15]. Ein Polystore ermöglicht es, Daten aus heterogenen Datenquellen in einer einzigen Anfrage miteinander zu verknüpfen.

Auch wenn dadurch die prinzipielle Auswertbarkeit dieser Daten gegeben ist, müssen die verschiedene Datenquellen immer noch weitgehend manuell von Datenanalysten zusammengeführt werden. Die Fragestellungen, die von den Datenanalysten beantwortet werden sollen, beziehen sich naturgemäß auf deren mentales Modell der Problemdomäne. Dieses sich ständig weiterentwickelnde Modell beinhaltet Informationen darüber, wann Datenquellen nützlich sind, wie diese verknüpft werden können, wie deren Inhalt zu interpretieren ist oder welches Vokabular verwendet wird. Derartige Wissensaspekte werden vielfach nicht dokumentiert, sondern unter Kollegen informell weitergegeben. Besonders in größeren Organisationen arbeiten häufig mehrere Teams von Datenanalysten an der Beantwortung unterschiedlicher Fragestellungen. Wenn diese Teams nicht miteinander interagieren, verpassen sie wichtige Gelegenheiten, ihr Wissen über den Datenbestand zu teilen oder von den Erfahrungen anderer zu profitieren.

Aus diesem Grund schlagen wir die Entwicklung von *Anfrage-getriebenen Wissenstransfersystemen (AWTS)* vor. Ein AWTS erweitert ein Datenmanagementsystem, beispielsweise einen Polystore, um neuartige Dienste, die Wissen aus Anfragen formalisieren und allen Benutzern zur Verfügung stellen können. Dabei wird ausgenutzt, dass das einer Anfrage zugrundeliegende mentale Modell partiell extrahiert und analysierbar gemacht werden kann. Das System unterstützt die Abbildung dieses Modells auf den tatsächlichen Datenbestand, indem die bisherigen Abbildungen anderer, ähnlicher Anfragen wiederverwendet werden.

## 2   Anfrage-getriebener Wissenstransfer

Die Schaffung von neuem Wissen spielt für den Alltag von Datenanalysten eine große Rolle. Im Folgenden bezeichnen wir mit dem Begriff *Wissen* Domänenwissen über verfügbare Datenquellen, das benötigt wird, um die Quellen für Analysezwecke zu verwenden. Dieses Wissen umfasst unter anderem folgende Aspekte: **(1)** Welche Datenquellen sind verfügbar? **(2)** Welche Teile dieser Datenquellen sind für welche Zwecke geeignet? **(3)** Welches Vokabular und welche Semantik wird zur Beschreibung des Inhalts bestimmter Datenquellen verwendet? **(4)** Wie können Datenquellen miteinander verknüpft werden? **(5)** Welche relevanten Informationen können aus dem Inhalt der Datenquellen abgeleitet werden? **(6)** Wer verwendet welche Datenquellen in welchem zeitlichen Kontext?

Wir beziehen uns auf das zirkuläre SEKI-Modell von Ikujiro Nonaka [NK98], um zu erläutern, wie ein AWTS die Schaffung von neuem Wissen unterstützen kann (Abb. 1). Das Modell unterscheidet zwischen *explizitem* und *stillem* Wissen. Ersteres kann direkt formuliert und geteilt werden, wohingegen letzteres persönliche Fähigkeiten und mentale Modelle umfasst, die nicht direkt formalisiert werden. Stilles Wissen wird durch *Sozialisierung* weitergegeben. Dieses Wissen wird durch *Externalisierung* explizit gemacht, beispielsweise durch textuelle Dokumentation. Durch *Kombination* wird neues Wissen durch Analyse des verfügbaren expliziten Wissens geschaffen. Bei der *Internalisierung* wird explizites Wissen von Individuen verinnerlicht und in stilles Wissen umgewandelt.

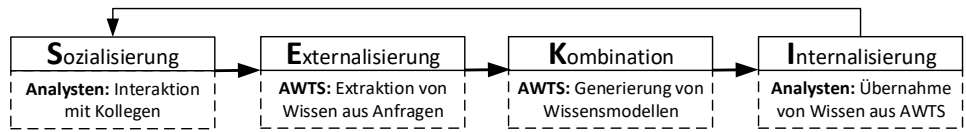

Abb. 1: SEKI-Modell am Beispiel von anfrage-getriebenem Wissenstransfer

Datenanalysten wenden ihre Erfahrungen und mentalen Modelle an, um analytische Fragestellungen zu formulieren. Stilles Wissen wird bereits durch Interaktion innerhalb von Analystenteams geteilt. Ein AWTS fördert die Externalisierung dieses Wissens, indem alle analytischen Anfragen innerhalb einer Organisation gesammelt werden. Teile der mentalen Modelle der Analysten werden somit in strukturierter Form erfasst und können explizit gemacht werden. Das AWTS übernimmt die Kombination dieses Wissens durch die Generierung von Wissensmodellen, die Muster, Beziehungen und Hinweise zur Semantik von Datenquellen aus den Anfragen ableiten. Auf Basis dieser Modelle erhalten die Analysten Empfehlungen über Datenquellen. Zur Internalisierung dieses neuen Wissens können die Empfehlungen analysiert und in zukünftigen Anfragen berücksichtigt werden.

### 2.1  Vorteile der Verwendung eines AWTS aus Sicht der Datenanalysten

Um die Vorteile eines AWTS für Datenanalysten zu verdeutlichen, erläutern wir dessen Verwendung anhand eines praxisnahen Szenarios aus der klinischen Forschung. Wir betrachten drei Teams von Datenanalysten, die ein zentralisiertes AWTS verwenden (Abb. 2). Das AWTS verwaltet verschiedene Daten aus elektronischen Krankenakten, die zusätzlich für die klinische Forschung genutzt werden sollen („secondary use"[1]). Zum besseren Verständnis werden alle Beispielanfragen stark vereinfacht dargestellt.

Team 1 verwendet Medikationspläne aus der Datenquelle D1 (Speicherformat JSON) und anonymisierte Patientendaten aus D2 (relationale Speicherung), um die Dosierung von Medikamenten zu analysieren. Team 2 führt ebenfalls Analysen der Medikationsgabe durch, verlässt sich dabei aber neben D2 auf die Medikationspläne aus D3 (Speicherformat CSV) sowie D4 (relationale Speicherung). Team 3 ist auf die Zeitreihenanalyse der Messwerte von Patientenmonitoren spezialisiert. Sie verwendet dazu die Datenquellen D5 (Speicherformat Avro) und D6 (Speicherformat Parquet) aus einem verteilten Dateisystem. Initial wissen die Teams nicht voneinander. Das AWTS bietet eine einheitliche Anfrageschnittstelle für verschiedene Arten von Datenquellen (im Beispiel SQL). Anfragen werden beim Zugriff auf die Datenquellen entsprechend übersetzt.

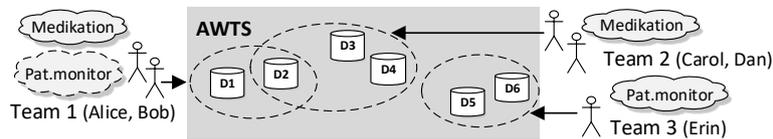

Abb. 2: Drei Teams von Datenanalysten interagieren mit einem AWTS

Während Alice von Team 1 auf die Ergebnisse einer Anfrage (Abb. 3) wartet, erkennt das AWTS auf Basis bestimmter Anfragen von Team 2 (Abb. 4), dass beide Teams die Datenquelle D2 verwenden. Über eine graphische Schnittstelle stellt das AWTS Alice nun Informationen und Empfehlungen für zukünftige Anfragen auf Basis des Wissens von Team 2 bereit. Durch Analyse der Anfragen von Team 2 erkennt das AWTS, dass D3 und D4 bereits mit D2 verknüpft worden sind. Daher präsentiert es Alice direkt eine einheitliche Sicht auf diese Datenquellen. Um Alice bei der Analyse von D3 und D4 zu unterstützen, gibt das AWTS unter anderem Auskunft über die Verwendungshäufigkeit und den Verwendungskontext in den bisherigen Analysesitzungen von Team 2.

```
SELECT D2.id, D2.Abteilung
FROM D1 JOIN D2 ON D2.id = D1.PatNr
WHERE D1.Wirkstoff LIKE 'Dexametha%';
```

Abb. 3: Anfrage von Alice (Team 1)

```
SELECT MIN(D2.Alter)
FROM D4 JOIN D2 USING id
WHERE D4.Arzneistoff = 'Salbutamol';
```
```
SELECT D2.id, D3.Substance, D3.Dose
FROM D3 JOIN D2 ON D2.id = D3.Patient;
```

Abb. 4: Anfrageprotokollauszug von Team 2

---

[1] Vgl. zu diesem Thema auch Arbeiten der zugeh. AG der GMDS: http://www.pg-ss.imi.uni-erlangen.de/

Bob von Team 1 hat kürzlich die Aufgabe erhalten herauszufinden, inwiefern die Gabe einiger Wirkstoffe mit dem Auftreten kritischer Werte bestimmter Vitalparametern korreliert. Er kennt noch keine Datenquellen, die diese Parameter enthalten, aber er hat eine ungefähre Vorstellung von den Daten, die er sucht. Er verwendet sein mentales Modell, um eine Anfrage zu schreiben, welche die hypothetische Datenquelle `Vitalparameter` referenziert (Abb. 5). Bob geht davon aus, dass die gesuchte Datenquelle die Attribute `Herzfrequenz` und `Blutdruck` besitzt. Das AWTS verwendet seine Annahmen, um geeignete Datenquellen zu empfehlen. Durch Auswertung der Struktur der im Anfrageprotokoll referenzierten Datenquellen erkennt das AWTS, dass Erin von Team 3 schon einmal Datenquellen verwendet hat, die möglicherweise für Bob geeignet sind (Abb. 6). Daher schlägt das AWTS die Verwendung von D5 (Attribute `HeartRate`, `BloodPressure`, `Time`, `Patient`) und D6 (Attribute `HRt`, `BTemp`, `BP`, `TStmp`, `PId`) vor. Bob kann nun entscheiden, ob er den Empfehlungen folgen will. Wenn Bob dies tut, wird seine Anfrage vom AWTS derart modifiziert, dass an Stelle von `Vitalparameter` mindestens eine der gefundenen Datenquellen referenziert wird. Die von Erin verwendeten Umbenennungen werden zudem Teil einer Ontologie, mit deren Hilfe Bob das Vokabular anderer Teams einfacher erfassen kann. Über einen interaktiven Dialog kann er eine Rückmeldung bezüglich der Vorschläge an das AWTS übermitteln. Das AWTS speichert die Abbildung zwischen seinem mentalen Modell und den tatsächlich verfügbaren Datenquellen und erzeugt automatisch eine individuelle Sicht, die mit seinem mentalen Modell korrespondiert. Bob kann diese Sicht direkt in zukünftigen Anfragen verwenden, ohne dass zusätzliche Interaktionen mit dem AWTS notwendig sind.

```
SELECT Herzfrequenz, Blutdruck
FROM Vitalparemeter
WHERE Herzfrequenz >= 130;
```

Abb. 5: Anfrage von Bob (Team 1)

```
SELECT Patient, BloodPressure AS Blutdruck
FROM D5
WHERE HeartRate < 90 AND Time > '16-01-10';
```
```
SELECT HRt AS Herzfrequenz, BP AS Blutdruck
FROM D6
WHERE PId = 'P41' AND TStmp = '1475693932';
```

Abb. 6: Anfrageprotokollauszug von Erin (Team 3)

Sobald Synergien zwischen verschiedenen Teams identifiziert werden, können die Datenanalysten das Wissen anderer direkt in die formalisierte Repräsentation ihrer mentalen Modelle übernehmen. Das AWTS unterstützt somit eine vereinfachte Datenintegration. Zum Beispiel kann Bob das Wissen von Erin nutzen, solange er an Analysen von Vitalparametern beteiligt ist. Das bedeutet, dass er dieselbe Sicht auf die Vitalparameter erhält wie Erin selbst. Wenn er sich nicht mehr für diese Daten interessiert, kann er auch wieder zu seiner auf Medikationspläne fokussierten Sicht des Datenbestands zurückkehren.

### 2.2    Formalisierung von Wissen aus analytischen Anfragen

Um die beschriebenen Dienste des AWTS zu ermöglichen, führen wir *Wissensfragmente* als Abstraktionskonzept für das kollektive Wissen aus dem Anfrageprotokoll ein. Ein Fragment formalisiert das mentale Modell einer Gruppe von Datenanalysten zu den verfügbaren Datenquellen über einen bestimmten Zeitraum. Wir verwenden die Bezeichnung *Wissensfragment*, da individuelle mentale Modelle unvollständig sein können, wohingegen die Kombination aller mentalen Modelle in einer Organisation das gesamte Wissen über den

Datenbestand bildet. Wie in Abb. 7 zu sehen ist, kapseln die Fragmente verschiedene Wissensmodelle, die wiederum auf Ausschnitten des Anfrageprotokolls basieren. Nachfolgend beschreiben wir mit Hilfe eines formalen Modells den Lebenszyklus von Fragmenten und erläutern ihre Verwendung durch Datenanalysten.

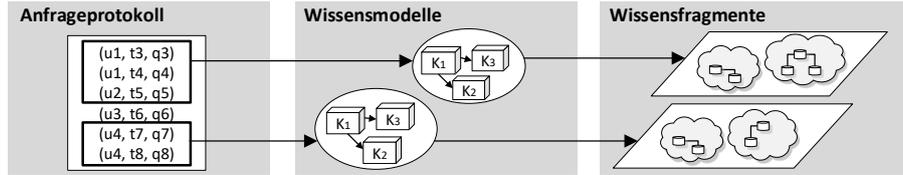

Abb. 7: Erzeugung von Wissensfragmenten aus dem Anfrageprotokoll

### 2.2.1 Erzeugung und Lebenszyklus von Wissensfragmenten

Wissensfragmente werden gemäß der Wünsche der Datenanalysten erzeugt. Sie bestimmen, welches Wissen in Form von Anfragen berücksichtigt werden soll.
→ **Beispiel:** Initial ist ein Fragment pro Analystenteam sinnvoll, das alle bisherigen Anfragen der Beteiligten beinhaltet. Im Beispielszenario aus Abschnitt 2.1 verwaltet das AWTS zu Beginn daher drei Fragmente.

**Extraktion relevanter Anfragen aus dem Anfrageprotokoll** Wir modellieren das Anfrageprotokoll $L$ als Menge von Einträgen vom Typ $\mathcal{L}$. Jeder Logeintrag umfasst einen Benutzerschlüssel $u$ vom Typ $\mathcal{U}$, einen Zeitstempel $t$ vom Typ $\mathcal{T}$ und eine Anfrage $q$ vom Typ $\mathcal{Q}$ (Abb. 8 (1)). Nur bestimmte Teile des Anfrageprotokolls $L$ sind für die Datenanalysten interessant. Diese Teile werden mit Funktionen extrahiert, die durch Anwendung einer Menge von Filterprädikaten vom Typ $\mathcal{F}$ auf $L$ eine Teilmenge des Anfrageprotokolls zurückliefern (Abb. 8 (2)). Dabei sind viele Arten von Filterprädikaten und beliebige Extraktionsfunktionen denkbar. Wir stellen exemplarische Definitionen bereit, die für ein AWTS geeignet sind: Jedes Filterprädikat beinhaltet den Benutzer $u$, dessen Anfragen extrahiert werden sollen, sowie einen Zeitraum für die Extraktion in Form der Zeitstempel $t_{start}$ und $t_{end}$ (Abb. 8 (3)). Unter Verwendung dieser Filterdefinition führen wir eine Funktion `extr` ein, die alle relevanten Anfragen aus dem Anfrageprotokoll $L$ extrahiert (Abb. 8 (4)).
→ **Beispiel:** Im Beispielszenario aus Abschnitt 2.1 werden initial die Filterprädikate $F_1$ bis $F_3$ zur Extraktion der relevanten Anfragen verwendet (Abb. 9). Dabei werden jeweils alle Anfragen aller Teammitglieder berücksichtigt. $t_\alpha$ und $t_\omega$ zeigen an, dass alle Anfragen über den gesamten Lebenszyklus des AWTS zu berücksichtigen sind. Selbstverständlich ist auch die Angabe fester Zeiträume möglich.

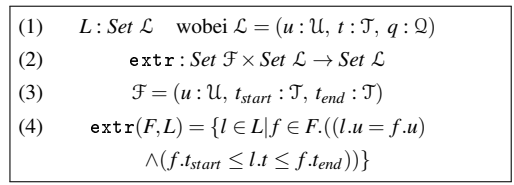

Abb. 8: Extraktion von relevanten Anfragen

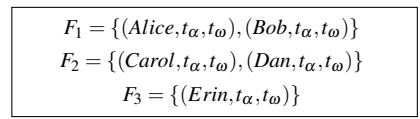

Abb. 9: Beispielhafte Filterprädikate

**Erzeugung von Wissensmodellen**   Aus den zuvor extrahierten Teilen des Anfrageprotokolls werden mit Hilfe verschiedener Algorithmen Wissensmodelle erzeugt, die interessante Aspekte der Anfragen herausarbeiten (Abb. 7). Diese Algorithmen $a_i$ erzeugen aus einer Teilmenge des Anfrageprotokolls individuelle Datenstrukturen $K_i$, um das extrahierte Wissen zu repräsentieren (Abb. 10 (5)). Die Indizes $i$ sind Elemente der Indexmenge $I$, mit deren Hilfe alle durch das AWTS bereitgestellten Algorithmen erfasst werden. Jedes Wissensmodell vom Typ $\mathcal{K}$ besteht aus dem Produkt der einzelnen, durch die Algorithmen extrahierten Wissensaspekte (Abb. 10 (6)). Die Funktion $\texttt{createModel}_{L,\texttt{extr},I}$ erzeugt ein Modell, indem die durch eine gegebene Indexmenge $I$ erfassten Algorithmen $a_i$ auf einen extrahierten Teil des Anfrageprotokolls angewendet werden (Abb. 10 (7)).

→ **Beispiel:** Um die im Beispielszenario aus Abschnitt 2.1 beschriebenen Funktionalitäten umzusetzen, bietet das AWTS verschiedene Algorithmen: So kann ein Algorithmus $a_{links}$ in Anfragen vorkommende Verbindungen zwischen Datenquellen in Form eines Graphen aufbereiten. Ein anderer Algorithmus $a_{session}$ kann das Anfrageprotokoll in Sitzungen segmentieren. Ein Algorithmus $a_{struct}$ analysiert die Struktur der angefragten Teile von Datenquellen. Um Ähnlichkeiten zwischen den Vokabularen verschiedener Teams zu erkennen, kann $a_{onto}$ eine Ontologie verwalten. Mithilfe der sich ergebenden Indexmenge $I$ (Abb. 11 (8)) kann das Wissensmodell $\mathcal{K}_1$ für Team 1 erzeugt werden (Abb. 11 (9)).

$$
\begin{aligned}
(5) \quad & a_i : \mathit{Set}\ \mathcal{L} \to K_i \quad \text{für } i \in I \\
(6) \quad & \mathcal{K} = \prod_{i \in I} K_i = (K_{i_1}, \dots, K_{i_n}) \\
(7) \quad & \texttt{createModel}_{L,\texttt{extr},I} : \mathit{Set}\ \mathcal{F} \to \mathcal{K} \\
& \texttt{createModel}_{L,\texttt{extr},I}(F) = \prod_{i \in I} a_i\,(\texttt{extr}(F,L))
\end{aligned}
$$

Abb. 10: Erzeugung von Wissensmodellen

$$
\begin{aligned}
(8) \quad & I = \{links, session, struct, onto\} \\
(9) \quad & \mathcal{K}_1 = \texttt{createModel}_{L,\texttt{extr},I}(F_1) = \\
& = (a_{links}\,(\texttt{extr}(F_1,L)), a_{session}\,(\texttt{extr}(F_1,L)), \\
& \quad a_{struct}\,(\texttt{extr}(F_1,L)), a_{onto}\,(\texttt{extr}(F_1,L))) = \\
& = (K_{links}, K_{session}, K_{struct}, K_{onto})
\end{aligned}
$$

Abb. 11: Beispielhaftes Wissensmodell

**Wissensfragmente**   Ein Fragment vom Typ $\mathcal{S}$ formalisiert das mentale Modell der kollaborierenden Datenanalysten (Abb. 12 (10)). Dazu wird eine Menge $F$ von Filterprädikaten von einem Analysten spezifiziert, um das Fragment mit den relevanten Anfragen zu initialisieren. Diese Anfragen werden zur Parametrisierung eines Wissensmodells $K$ verwendet. Die Funktion $\texttt{createShard}_{L,\texttt{extr},I}$ nimmt die Menge $F$ von Filterprädikaten vom Typ $\mathcal{F}$, um für ein gegebenes Anfrageprotokoll $L$, eine gegebene Extraktionsfunktion $\texttt{extr}$ und eine gegebene Indexmenge $I$ von Algorithmen das zugehörige Fragment zu erzeugen (Abb. 12 (11)). $L$, $\texttt{extr}$ und $I$ sind für alle Fragmente eines bestimmen AWTS identisch.

→ **Beispiel:** Im Beispielszenario aus Abschnitt 2.1 erzeugen Alice, Carol und Erin gemäß der genannten Filterprädikate jeweils ein Fragment für ihr jeweiliges Team (Abb. 13).

$$
\begin{aligned}
(10) \quad & \mathcal{S} = (F : \mathit{Set}\ \mathcal{F},\ K : \mathcal{K}) \\
(11) \quad & \texttt{createShard}_{L,\texttt{extr},I} : \mathit{Set}\ \mathcal{F} \to \mathcal{S} \\
& \texttt{createShard}_{L,\texttt{extr},I}(F) = (F, \texttt{createModel}_{L,\texttt{extr},I}(F))
\end{aligned}
$$

Abb. 12: Erzeugung von Wissensfragmenten

$$
\begin{aligned}
s_1 &= \texttt{createShard}_{L,\texttt{extr},I}(F_1) \\
s_2 &= \texttt{createShard}_{L,\texttt{extr},I}(F_2) \\
s_3 &= \texttt{createShard}_{L,\texttt{extr},I}(F_3)
\end{aligned}
$$

Abb. 13: Bsph. Wissensfragmente

**Strukturelle Operationen**   Wissensfragmente sind dynamisch und können sich im Verlauf der Zeit weiterentwickeln, indem neue Anfragen Teil der jeweiligen Wissensmodelle werden. Um den Datenanalysten mehr Flexibilität zu ermöglichen, stellt das AWTS verschiedene strukturelle Operationen bereit: Zwei Fragmente können verschmolzen werden, um eine weiterführende Kooperation zwischen zwei Analystenteams zu modellieren. Die Funktion $\text{merge}_{L,\text{extr},I}$ erzeugt dazu mit Hilfe der Filterprädikate zweier Fragmente ein neues Fragment (Abb. 14 (12)). Fragmente können mit $\text{expand}_{L,\text{extr},I}$ (Abb. 14 (13)) vergrößert oder mit $\text{narrow}_{L,\text{extr},I}$ (Abb. 14 (14)) verkleinert werden. Dadurch können entweder bestimmte Teile des Anfrageprotokolls hinzugefügt oder ausgeschlossen werden.

$$
\begin{aligned}
&(12) \quad \text{merge}_{L,\text{extr},I}: \mathcal{S} \times \mathcal{S} \to \mathcal{S} \quad \text{merge}_{L,\text{extr},I}(s1,s2) = \text{createShard}_{L,\text{extr},I}(s1.F \cup s2.F) \\
&(13) \quad \text{expand}_{L,\text{extr},I}: \mathcal{S} \times \mathcal{F} \to \mathcal{S} \quad \text{expand}_{L,\text{extr},I}(s,F) = \text{createShard}_{L,\text{extr},I}(s.F \cup F) \\
&(14) \quad \text{narrow}_{L,\text{extr},I}: \mathcal{S} \times \mathcal{F} \to \mathcal{S} \quad \text{narrow}_{L,\text{extr},I}(s,F) = \text{createShard}_{L,\text{extr},I}(s.F - F)
\end{aligned}
$$

Abb. 14: Basisoperationen auf Wissensfragmenten

**Vergleichsoperationen** Das Erkennen von Ähnlichkeiten zwischen Fragmenten ist ein wichtiger Faktor für deren Weiterentwicklung. Zwei Fragmente sind genau dann ähnlich, wenn ihre Wissensmodelle ähnlich sind. Für den Vergleich von Wissensmodellen wird auf Funktionen $\text{sim}K_i$ zurückgegriffen, mit denen jeweils von einem bestimmten Algorithmus $a_i$ aus dem Anfrageprotokoll extrahiertes Wissen verglichen werden kann (Abb. 15 (15)). Wissensmodelle vom Typ $\mathcal{K}$ werden mit Hilfe der Ähnlichkeitsfunktion $\text{sim}\mathcal{K}$ verglichen. Mit Hilfe einer Gewichtsfunktion $w$ kann der Einfluss einzelner Algorithmen aus der Indexmenge $I$ bestimmt werden (Abb. 15 (16)). Die Funktion $\text{sim}\mathcal{S}$ bestimmt die Ähnlichkeit zweier Fragmente durch den paarweisen Vergleich ihrer Wissensmodelle (Abb. 15 (17)) gemäß der Gewichtungsfunktion $w$. Auf Basis dieses Ähnlichkeitsmaßes generiert das AWTS unter anderem Empfehlungen für strukturelle Operationen.

→ **Beispiel:** Es wird vereinfachend angenommen, dass alle Algorithmen bis auf $a_{struct}$ die Gewichtung 0 erhalten. Aufgrund der strukturellen Ähnlichkeit der Fragmente $s_1$ und $s_2$ im Beispielszenario aus Abschnitt 2.1 ($s_2$ enthält die Hälfte aller Datenquellen von $s_1$, vgl. Abschnitt 2.1) kann das AWTS empfehlen $s_2$ mit $s_1$ zu verschmelzen.

$$
\begin{aligned}
&(15) \quad \text{sim}K_i : K_i \times K_i \to [0;1] \\
&(16) \quad \text{sim}\mathcal{K} : (I \to [0;1]) \times \mathcal{K} \times \mathcal{K} \to [0;1] \\
&\quad \text{sim}\mathcal{K}(w, \mathcal{K}_1, \mathcal{K}_2) = \sum_{i \in I} w(i) \cdot \text{sim}K_i(\mathcal{K}_1.K_i, \mathcal{K}_2.K_i) \quad \text{wobei} \sum_{i \in I} w(i) = 1 \\
&(17) \quad \text{sim}\mathcal{S} : (I \to [0;1]) \times \mathcal{S} \times \mathcal{S} \to [0;1] \\
&\quad \text{sim}\mathcal{S}(w, s_1, s_2) = \text{sim}\mathcal{K}(w, s_1.K, s_2.K)
\end{aligned}
$$

Abb. 15: Ähnlichkeitsfunktionen für Wissensmodelle und Wissensfragmente

### 2.2.2 Anfrageverarbeitung

Bei der Benutzung eines AWTS formulieren die Benutzer Anfragen nicht gegen den tatsächlichen Datenbestand, sondern gegen ihr formalisiertes mentales Modell (Abb. 16). Zur Anfrageverarbeitung wird jede Anfrage protokolliert. Das System entscheidet dann, ob die Anfrage nur tatsächlich vorhandene Schemaelemente enthält. Ist dies der Fall, wird die Anfrage normal verarbeitet und die Ergebnisse werden zurückgeliefert. Parallel dazu werden die relevanten Wissensmodelle durch eine Inferenzmaschine ausgewertet. Dadurch gewonnene Empfehlungen (z.B. ähnliche Datenquellen oder Benutzer mit ähnlichem Informationsbedarf) werden dem Benutzer in einer separaten Oberfläche angezeigt.

Damit die Analysten ihre mentalen Modelle in ihren Anfragen verwenden können, sind auch nicht tatsächlich vorkommende Schemaelemente in Anfragen zulässig. Dazu werden die Wissensmodelle ausgewertet, um die Anfragen durch Modifikationen verarbeitbar zu machen (z.B. Abbildung auf vorhandene Datenquellen). Bei Bedarf wird dazu eine Interaktion mit dem Benutzer angestoßen, sonst entscheidet das System autonom.

→ **Beispiel:** Für die Empfehlungen im Beispielszenario aus Abschnitt 2.1 greift das AWTS auf die Wissensmodelle der Fragmente $s_1$, $s_2$ und $s_3$ zurück.

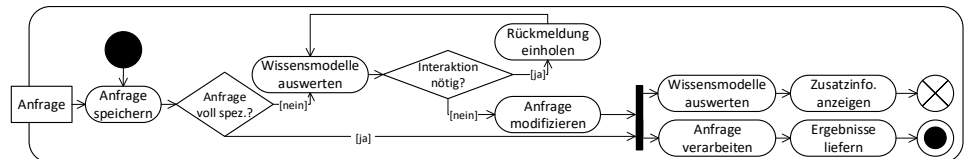

Abb. 16: Anfrageverarbeitung unter Einbezug von Wissensfragmenten

## 3  Referenzarchitektur

Für die Implementierung eines AWTS schlagen wir eine Client-Server-Architektur vor (Abb. 17). Ein AWTS kommuniziert mit einem vorhandenen Datenmanagementsystem (DMS), um Zugriff auf die verfügbaren Datenquellen zu ermöglichen. Benutzer formulieren Anfragen über die *programmatische Schnittstelle*, die die native Anfrageschnittstelle des DMS kapselt. Etablierte Analyseprozesse können beibehalten werden, da Analysewerkzeuge einfach mit dem AWTS statt mit dem DMS verbunden werden. Die *graphische Benutzeroberfläche* stellt für jede Anfrage relevantes Wissen dar und präsentiert Anfragemodifikationen. Darüber hinaus können hiermit Lebenszyklusoperationen auf den Fragmenten durchgeführt werden. Eingehende Anfragen werden protokolliert und nach eventuellen Modifikationen an das DMS weitergeleitet. Die *Fragmentverwaltung* ist für den Lebenszyklus aller Fragmente verantwortlich und erzeugt die Wissensmodelle. Zudem überwacht sie das Anfrageprotokoll, um bei Bedarf die Wissensmodelle zu aktualisieren. Die Komponente zum *Wissenstransfer* stellt relevantes Wissen bereit und stößt Anfragemodifikationen an. Dazu werden die Wissensmodelle mit Hilfe der *Inferenzmaschine* ausgewertet. Sie ist auch für die Überwachung der Fragmente zuständig, um Empfehlungen für Lebenszyklusoperationen abzuleiten.

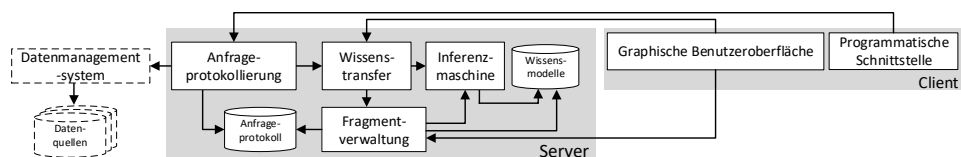

Abb. 17: Referenzarchitektur

Für weitere Untersuchungen implementieren wir aktuell auf Basis von *Apache Drill*[2] ein AWTS, das den Zugriff auf verschiedene Typen von Datenquellen mittels SQL ermöglicht. Dazu werden unter anderem Algorithmen für Anfrageähnlichkeit, Sitzungserkennung, Schemazusammenführung und Gewichtung von Verbundattributen bereitgestellt.

---

[2] https://drill.apache.org

## 4 Evaluationsansatz

Die Evaluation unserer Ergebnisse erfolgt zweistufig. Zum einen wird überprüft, ob unser Ansatz einen konkreten Mehrwert für Datenanalysten bietet und ihnen die Beantwortung analytischer Fragestellungen erleichtert. Zum anderen wird untersucht, ob die Leistungsfähigkeit unserer Referenzimplementierung für analytische ad-hoc-Anfragen ausreichend und somit eine interaktive Verwendung des Systems möglich ist.

Der praktische Nutzen unseres Ansatzes wird mit einer Benutzerstudie in einer konkreten Anwendungsdomäne (z.B. klinische Forschung) überprüft. In einem ersten Schritt wird dazu die Korrektheit der Wissensmodelle evaluiert. Auf Basis von realen Benutzeranfragen werden Modelle erzeugt. Die Benutzer bewerten anschließend, inwiefern diese Modelle ihre Intentionen widerspiegeln. Darüber hinaus werden die zur Modellerstellung notwendige Anzahl von Anfragen und die Güte der Ähnlichkeitsfunktionen zum Modellvergleich untersucht. In einem zweiten Schritt erhalten zwei Benutzergruppen die Aufgabe, konkrete Fragestellungen mit Hilfe vorgegebener Datenquellen zu beantworten. Beide Gruppen verwenden das AWTS, um auf die Daten über eine einheitliche Schnittstelle zuzugreifen. Aber nur eine Gruppe wird vom System durch Empfehlungen auf Basis von Wissensmodellen unterstützt. Durch einen Vergleich der Arbeitsgeschwindigkeit und der Ergebnisqualität beider Benutzergruppen wird untersucht, ob die Funktionen zum Anfrage-getriebenen Wissenstransfer die Beantwortung analytischer Fragestellungen erleichtern.

Zur Bestimmung der Leistungsfähigkeit unserer Referenzimplementierung werden Rechenzeit und Speicherverbrauch für die Modellerzeugung, die Modellaktualisierung, die Modellauswertung und den Modellvergleich gemessen. Zudem werden die Antwortzeiten des Gesamtsystems bei der Verarbeitung von ad-hoc-Anfragen untersucht.

Bei der Evaluation der Leistungsfähigkeit kann unter anderem auf synthetische Anfrageprotokolle zurückgegriffen werden, die von einem Anfragegenerator erzeugt werden. Zur Beurteilung der Nützlichkeit unseres Ansatzes sind reale Anfrageprotokolle erforderlich. Neben den Anfragen, die im Rahmen der Benutzerstudie anfallen, können zusätzlich auch Anfragen aus bereits eingesetzten Datenmanagementsystemen herangezogen werden.

## 5 Verwandte Arbeiten

Unser Ansatz ist mit Dataspace-Systemen [FHM05] verwandt, die auf Basis von Benutzerrückmeldungen die verwalteten Daten inkrementell an die Erwartungen der Benutzer anpassen. Existierende Systeme berücksichtigen jedoch keine Szenarien, in denen Benutzergruppen mit heterogenen Erwartungen mit einer gemeinsamen Menge von Datenquellen arbeiten. Explizite Benutzerrückmeldungen waren bereits Forschungsgegenstand, Wissen aus analytischen Anfragen bleibt jedoch noch ungenutzt [Be13]. Khoussainova et al. fordern die Entwicklung von Systemen zur Verwaltung von Anfragen, um die darin enthaltenen Informationen nutzbar zu machen [Kh09]. Teile ihrer Vorschläge wurden von verschiedenen Forschungsprojekten zur Auswertung einzelner Aspekte von Anfrageprotokollen aufgegriffen (für einen Überblick siehe unsere Vorarbeiten [Wa16, Sc16]).

Mentale Modelle von Analysten sind nicht immer kongruent zum vorhandenen Datenbestand. Damit sie ihre mentalen Modelle dennoch in Anfragen verwenden können, er-möglichen einige Ansätze die Referenzierung unbekannter Schemaelemente in Anfragen [Eb15, LPJ14]. Komplexe temporale und soziale Zusammenhänge, die aus dem Anfrageprotokoll extrahiert werden können, finden jedoch keine Berücksichtigung.

Unser Ansatz unterscheidet sich von der Empfehlung ganzer Anfragen [Ei14] und der automatischen Vervollständigung von Anfragen [Kh10], weil wir den Analysten das Formulieren vollständiger Anfragen unter Verwendung ihrer mentalen Modelle ermöglichen. Ziel ist es, den Datenbestand an die mentalen Modelle der Benutzer anzupassen und nicht die Benutzer zur Anpassung ihrer mentalen Modelle zu zwingen.

## 6  Zusammenfassung

Um von der Verfügbarkeit von heterogenen Datenquellen zu profitieren, muss deren effiziente Nutzung sichergestellt werden. Wir schlagen *Anfrage-getriebene Wissenstransfersysteme (AWTS)* als Ansatz zur inkrementellen Datenintegration vor. Mit einem AWTS können Datenanalysten stilles Wissen über Datenquellen ohne manuellen Dokumentationsaufwand externalisieren. Sie lernen, wie andere Analysten die Datenquellen verwenden und können die Semantik relevanter Datenquellen einfacher verstehen. Ein AWTS stellt neuartige Funktionalitäten bereit, die Datenanalyseprozesse unterstützen können. Gemäß unserer Referenzarchitektur erfordert der Einsatz eines AWTS keine Änderung von etablierten Analyseprozessen, da bestehende Analysewerkzeuge weiterhin verwendet werden können.

## Literatur

[Be13]  Belhajjame, K.; Paton, N. W.; Embury, S. M.; Fernandes, A. A.; Hedeler, C.: Incrementally Improving Dataspaces Based on User Feedback. Inf. Syst., 38(5):656–687, Juli 2013.

[Eb15]  Eberius, J.; Thiele, M.; Braunschweig, K.; Lehner, W.: DrillBeyond: Processing Multi-result Open World SQL Queries. In: SSDBM '15. ACM, S. 16:1–16:12, 2015.

[Ei14]  Eirinaki, M.; Abraham, S.; Polyzotis, N.; Shaikh, N.: QueRIE: Collaborative Database Exploration. IEEE Trans. on Knowledge and Data Eng., 26(7):1778–1790, Juli 2014.

[FHM05]  Franklin, M.; Halevy, A.; Maier, D.: From Databases to Dataspaces: A New Abstraction for Information Management. SIGMOD Rec., 34(4):27–33, Dezember 2005.

[Kh09]  Khoussainova, N.; Balazinska, M.; Gatterbauer, W.; Kwon, Y.; Suciu, D.: A case for a collaborative query management system. arXiv preprint arXiv:0909.1778, 2009.

[Kh10]  Khoussainova, N.; Kwon, Y.; Balazinska, M.; Suciu, D.: SnipSuggest: Context-aware Autocompletion for SQL. Proc. VLDB Endow., 4(1):22–33, Oktober 2010.

[LPJ14]  Li, F.; Pan, T.; Jagadish, H. V.: Schema-free SQL. In: SIGMOD'14. ACM, S. 1051–1062, 2014.

[NK98]  Nonaka, I; Konno, N.: The concept of "ba": Building a foundation for knowledge creation. California mgmt. review, 40(3):40–54, 1998.

[Sa11]  Saeed, M.; Villarroel, M.; Reisner, A. T.; Clifford, G.; Lehman, L. et al.: Multiparameter Intelligent Monitoring in Intensive Care II (MIMIC-II): a public-access intensive care unit database. Critical care medicine, 39(5):952, 2011.

[Sc16]  Schwab, P. K.; Wahl, A. M; Lenz, R.; Meyer-Wegener, K.: Query-driven Data Integration (Short Paper). In: FG-DB'16. GI, 2016.

[St15]  Stonebraker, M.: The Case for Polystores. Juli 2015. http://wp.sigmod.org/?p=1629, abgerufen: 01.09.2016.

[Wa16]  Wahl, A. M.: A minimally-intrusive approach for query-driven data integration systems. In: ICDEW'16. IEEE, S. 231–235, 2016.